\documentclass{article}

\usepackage{amsmath,amsfonts,amsthm,amssymb}

\usepackage{authblk}

\usepackage{multicol}

\usepackage[pdftex]{graphicx}

\usepackage{url}

\usepackage{tabulary}
\usepackage{braket}

\usepackage{listings}
\usepackage{xcolor}
\usepackage{algorithm}
\usepackage{algpseudocode}

\usepackage{indentfirst}

\usepackage[margin={2.0cm,3.5cm}]{geometry}

\begin{document}

\onecolumn{
\title{\vspace{-0.8in} Massively parallel implementation and approaches to simulate quantum dynamics using Krylov subspace techniques}
\author{\large{Marlon Brenes,$^1$ Vipin Kerala Varma,$^{1,2,3,4}$ Antonello Scardicchio,$^{1,5}$ and Ivan Girotto$^{1,}$$^6$}\\
\normalsize{$^1$The Abdus Salam ICTP, Strada Costiera 11, 34151 Trieste, Italy\\
$^2$ Initiative for the Theoretical Sciences, The Graduate Center, CUNY, New York, NY 10016, USA \\
$^3$ Department of Engineering Science and Physics, College of Staten Island, CUNY, Staten Island, NY 10314, USA\\
$^4$ Department of Physics and Astronomy, University of Pittsburgh, Pittsburgh, PA 15260, USA\\
$^5$ INFN, Sezione di Trieste, Via Valerio 2, 34127, Trieste, Italy\\
$^6$University of Modena and Reggio Emilia, 41121 Modena, Italy}
}
\date{}

\maketitle
\vspace{-0.3in}
\begin{abstract}
We have developed an application and implemented parallel algorithms in order to provide a computational framework suitable for massively parallel supercomputers to study the unitary dynamics of quantum systems. We use renowned parallel libraries such as PETSc/SLEPc combined with high-performance computing approaches in order to overcome the large memory requirements to be able to study systems 
whose Hilbert space dimension comprises over 9 billion independent quantum states. Moreover, we provide descriptions on the parallel approach used for the three most important stages of the simulation: 
handling the Hilbert subspace basis, constructing a matrix representation for a generic Hamiltonian operator and the time evolution of the system by means of the Krylov subspace methods. 
We employ our setup to study the evolution of quasidisordered and clean many-body systems, focussing on the return probability and related dynamical exponents: the large system sizes accessible 
provide novel insights into their thermalization properties.

\end{abstract}
}

\begin{multicols}{2}

\section{Introduction} 

In the last 20 years the study of how quantum systems may or may not reach equilibrium has received a renovated interest. This is due mainly to experimental progress in maintaining mesoscopic quantum systems 
isolated from the environment (i.e.\ from decoherence) be they cold atoms in optical traps \cite{jaksch1998cold}, arrays of superconducting qubits \cite{steffen2006measurement} or impurities in crystals 
\cite{maze2008nanoscale}. During the early years of quantum mechanics, it was established that equilibrium states of ergodic systems \cite{goldstein2010normal} must be effectively described with proper quantum 
statistical mechanics models. The modern view building on those early works is contained in the so-called ``eigenstate thermalization hypothesis" \cite{PhysRevA.43.2046}. 
However, the precise mechanism of how these states can be reached by local dynamics that follow microscopic laws is yet to be understood (see \cite{eisert, cazadilla} for a review) and counterexamples to widely 
held conjectures about thermalization are being found, for example when quenched disorder is present in the system. 
Systems with quenched disorder might be constructed (and are indeed quite natural) to effectively behave like integrable systems 
\cite{basko2006metal,nandkishore2015many,serbyn2013local,ros2015integrals,imbrie2016review}, with their time evolution constrained to the conservation of extensively many local observables. 
These counterexamples have spurred a sort of dissonance between a microscopic description based on the Schr\"odinger equation of quantum mechanics and one that employs the classical ensembles 
from statistical mechanics, and have raised many conceptual questions, for example what are the nature and properties of such dynamical quantum phase transitions between thermalizing and non-thermalizing phases 
of matter. 

Lacking a quantum computer, simulating unitary time evolution of quantum dynamical systems on a classical computer is $-$ both in conceptual and computational terms $-$ a very demanding task. 
Given that studying quantum many-body systems out of equilibrium allows us to probe questions on the foundation of statistical mechanics and condensed matter theory, we provide one such quantum simulator 
that may be employed for theoretically analyzing the efficacy of quantum computing as well as other quantum technologies; 
a platform to perform numerical operations and simulations involving these systems is indeed very important to current research \cite{cazadilla}.

As rapid growth of entanglement is the main obstacle to simulating a quantum system on a classical computer, there are few techniques for calculating the time evolution of a quantum systems, mainly of two kinds. 
The first kind assumes that entanglement will be small during the whole evolution and uses an approximate form of the wave function. 
These algorithms fall under the umbrella of density-matrix renormalization group evolution or tDMRG (see \cite{schollwock2006methods} for a review). 
The second kind does not make this assumption and handles, from the very beginning, the largest possible wave function which can fit given the computational resources. 
The first kind of algorithms is suited for evolving states close to the ground state. One is practically guaranteed that, for most models, the entanglement between a subregion $A$ and a subregion $B$ of the system 
is bounded by the number of points at the boundary of $A$ and $B$. 
However, as far as the aforementioned thermalization properties are concerned, staying away from the ground state of the system is more pertinent and, for such questions, the tDMRG method will quickly fail.

One is then led to consider the second category of algorithms, which store the entire wave function in memory without making any assumption on its entanglement structure to begin with. 
A typical approach in this category is then to perform full diagonalization of the Hamiltonian, to obtain eigenvalues and eigenvectors and use these to evolve the initial state up to a given time $t$. 
However, for relatively large systems this practice is computationally problematic when it comes to actual computing times and memory requirements. 
A way out is the method of Krylov subspace techniques, of which a massively parallelized, effective implementation is the argument of this paper.

This paper is arranged as follows: Section 2 provides a brief background on the physics of the problem; we describe the implementation and approaches in Section 3; 
Section 4 comprises the Results and analysis, while Section 5 is devoted to Performance analysis of our implementation.   

\vspace{-0.1in}

\section{Background}

\begin{figure*}[htb]
\centering
\includegraphics[scale=0.9]{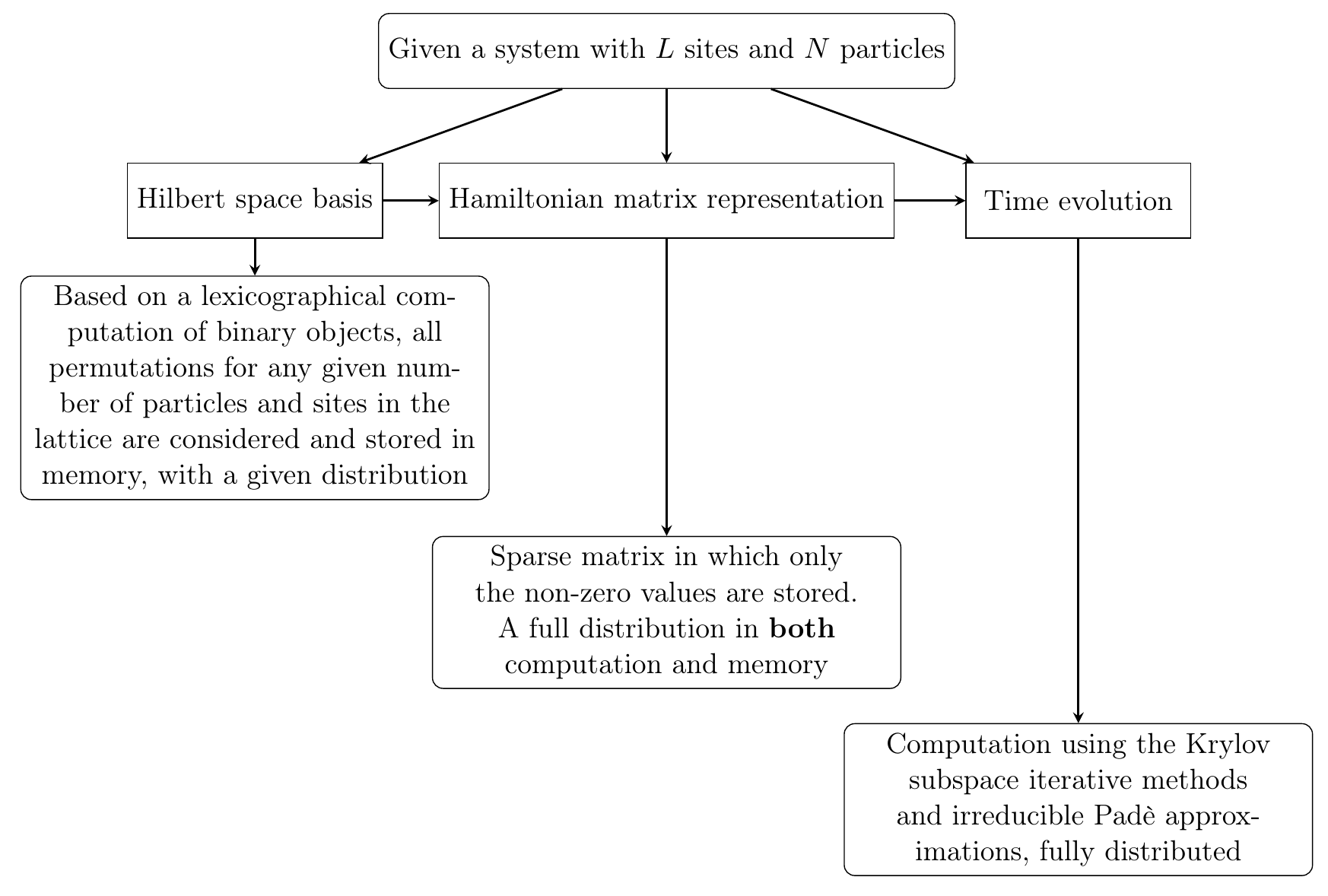} 
\caption{Brief summary of the parallel design for construction of the Hamiltonian and implementation of time-evolution of the system.}
\label{fig:design} 
\end{figure*}

In order to solve the quantum $N$ body problem in the framework of a quantum lattice system, sparse matrix algorithms are usually applied to solve for its corresponding properties and (eigen)states. 
A transformation of the considered many-particle Hamiltonian needs to be performed, in the language of the second quantization, into a sparse Hermitian matrix \cite{fehske}. 
Diverse models are under study to further our understanding of such slow or non-thermalizing systems, such as the many body localization phenomenon \cite{varma1,varma2} 
and to evaluate real-time dynamics of lattice gauge theories \cite{marcello}, to name a few; 
here such a transformation to sparse Hermitian matrices is common to study their dynamic properties.

{\em Models.} --- We focus our attention on a paradigmatic model for studying transport in one-dimensional system to illustrate the procedure of the parallel algorithms and implementation; 
namely, a system of one-dimensional {\em hardcore} bosons model with nearest neighbor hopping $t$ and nearest neighbor repulsion $V$ and an onsite quasidisordered potential of strength $h$, described by
\begin{align}
\label{eq: Hamiltonian}
\hat{H} &= t \sum_{i = 1}^{L-1}(\hat{c}_{i}^{\dagger}\hat{c}_{i+1} + \hat{c}_{i+1}^{\dagger}\hat{c}_{i}) \nonumber \\ &+ V \sum_{i = 1}^{L-1}(\hat{n}_{i}\hat{n}_{i+1}) + h \sum_{i = 1}^{L-1}(\hat{n}_{i}\cos({2\pi\beta i} + \phi))
\end{align}

where $\hat{c}_{i}^{\dagger}$ is the bosonic creation operator, $\hat{c}_{i}$ is the bosonic destruction operator and $\hat{n}_{i}$ is the bosonic counting operator such that 
$\hat{n}_{i} \equiv \hat{c}_{i}^{\dagger}\hat{c}_{i}$; $\beta$ is the inverse of the golden ratio given by $\frac{\sqrt{5} - 1}{2}$, $\phi$ is an arbitrary phase, and $L$ is the linear size of the physical chain.  
In this work we restrict to lattices with sizes corresponding to Fibonacci numbers ($L = F_{n+1}$) so that the disorder term completes a cycle along the length of the chain if $\beta$ is chosen as 
the ratio of successive Fibonacci numbers i.e. $\beta = \cfrac{F_n}{F_{n+1}}$.\\

The two main limits of the above Hamiltonian that we consider are as follows: 
(i) The XXZ Hamiltonian with $h=0, V \neq 0$ is a canonical integrable model for investigating transport and conservation laws in one-dimensional systems, with $V=0$ corresponding to the case of free particles; 
(ii) The Aubry-Andr\'e model with $V=0, h\neq 0$ is a {\em quasidisordered} model \cite{AubryAndre} which shows an Anderson delocalization-localization transition at all energy scales when $h=2$.

In such a system the operation $\hat{H}\ket{\Psi}$ returns a linear combination of other states in the Hilbert space basis, and so a Hamiltonian matrix can be constructed for the specific operator at hand. 
The result is a (usually very large) sparse Hermitian matrix. In this framework, an initial state can be prepared in order to study the system's unitary time evolution described by the Schr\"odinger 
equation:

\begin{align}
\label{eq: Schrodinger}
i\hbar\frac{\partial}{\partial t}|\Psi(\textbf{r},t)\rangle = \hat{H}|\Psi(\textbf{r},t)\rangle
\end{align}

with the corresponding solutions given by

\begin{align}
\ket{\Psi(t)} = e^{-i\hat{H}t/\hbar}\ket{\Psi(t=0)}
\end{align} 

From here on we set the Planck's constant $\hbar = 1$, and the spatial index $\textbf{r}$ is suppressed.

{\em Basis representation.} --- The full Hilbert space of the system comprises $2^L$ states corresponding to the presence or absence of a boson on each site. 
The Hamiltonian in Equation \eqref{eq: Hamiltonian} is $U(1)$ symmetric, thereby allowing us to focus on individual subspaces which are not connected by the Hamiltonian.
A proper representation of the basis vectors of the Hilbert subspace of dimension $\mathcal{D}$ needs to be devised in order to perform operations on the computer and to create 
a matrix representation of the Hamiltonian operator. In particular, for the case of the hardcore bosons described before, the dimension of each of these subspaces is given by 
$\mathcal{D} = L! / N! (L-N)!$, where $N$ denotes the number of bosonic particles.

A common approach that can be used consists in assigning an integer value to each of the basis states of the subspace. 
In this representation, each of the states correspond to a value in a memory buffer that can be traversed by lookup algorithms. For instance, for the case of $L=4$ and $N=2$, the states are represented by

\begin{equation}
	\begin{split}
	\ket{0011} &\rightarrow 3\\
	\ket{0101} &\rightarrow 5\\
	\ket{0110} &\rightarrow 6\\
	\ket{1001} &\rightarrow 9\\
	\ket{1010} &\rightarrow 10\\
	\ket{1100} &\rightarrow 12,\\
	\end{split} 
\end{equation}

where $0 (1)$ indicates the absence (presence) of a boson on that site.

\begin{algorithm*}
\caption{Parallel distribution}\label{Distribution}

\begin{algorithmic}

\Procedure{Distribution}{PetscInt \&$nlocal$, PetscInt \&$start$, PetscInt \&$end$}

  \State $nlocal$ = $basis\_size\_$ / $mpisize\_$
  \State PetscInt $rest$ = $basis\_size\_$ \% $mpisize\_$
  
  \If{$rest$ \&\& $(mpirank\_ < rest)$}
    \State $nlocal++$
  \EndIf
  
  \State  $start$ = $mpirank\_$ * $nlocal$
  
  \If{$rest$ \&\& $(mpirank\_ >= rest)$}
    \State $start$ += $rest$
  \EndIf
  
  \State $end$ = $start$ + $nlocal$
  
\EndProcedure

\end{algorithmic}

\end{algorithm*}

This is a very useful scheme to represent the basis states, given that integer values are easier to work with than binary objects. In particular, if the basis is stored as a contiguous memory buffer effective 
lookup algorithms can be used to search for a specific element; even more so if the elements are sorted . For instance, a binary lookup could be used to search for an element of an array of size $N$ with complexity $\mathcal{O}(\log N)$, compared to the complexity of $\mathcal{O}(N)$ 
for an element-by-element lookup.

{\em Krylov subspace methods.} --- Applying the methodology described, the problem translates to evaluate the exponential of a large sparse matrix for the system. We can apply the technique of Krylov 
subspaces in order to \textbf{avoid} full diagonalization. With this approach we approximate the solution to Equation \eqref{eq: Schrodinger} using power series. The optimal polynomial approximation to $\ket{\Psi(t)}$ from within 
the Krylov subspace (denoting $|\Psi_0 \rangle \equiv |\Psi(t=0)\rangle$),

\begin{align}
\mathcal{K}_m = &\textrm{span}\left\{\ket{\Psi_0}, H\ket{\Psi_0}, H^2\ket{\Psi_0}\dots,H^{m-1}\ket{\Psi_0}\right\},
\end{align} 

is obtained by an Arnoldi decomposition of the matrix $A_m = V_m^THV_m$ where $m$ is dimension of the Krylov subspace, such that $m \ll \mathcal{D}$. 
That is in order to attain convergence, only a much smaller dimension of the Krylov subspace is required in comparison to the dimension of the Hilbert subspace \cite{expokit}.\\
The desired time-evolved solution is then well-approximated by

\begin{align}
\ket{\Psi(t)}\approx V_{m}exp(-itA_m)\ket{e_1},
\end{align}
 
where $\ket{e_1}$ is the first unit vector of the Krylov subspace. 
The much smaller matrix exponential is then evaluated using irreducible Pad\`e approximations \cite{varma1}. It has been established that this approximation is better than the $m$-fold Taylor expansion and it has been shown \cite{expokit} that the error in the Krylov method behaves like $\mathcal{O}(e^{m-t||A||_{2}}(t||A||_{2}/m)^m)$ when $m \leq 2t||A||_{2}$, which indicates that the technique can work quite well even for moderate $m$ if a time-stepping strategy is embedded in the process along with error estimations. The algorithm to evaluate the numerical method has been extensively described by \cite{expokit}.

\section{Implementation}

\begin{algorithm*}
\caption{Next permutation of bits}\label{Perm}

\begin{algorithmic}[1]

\Procedure{ConstructIntBasis}{LLInt *$int\_basis$}

  \State LLInt $w$ \Comment{Next permutation of bits for construction of the basis}
  \State LLInt $smallest$ = $smallest\_int$() \Comment{Smallest int of the basis}
  \State $int\_basis[0]$ = $smallest$
  \For{LLInt $i = 1$ to $basis\_size\_ - 1$}
    \State LLInt $t$ = $($$smallest$ $|$ $($$smallest$ $- 1$$))$ $+ 1$
    \State $w$ = $t$ $|$ $(((($$t$ \& -$t$ / $($$smallest$ \& -$smallest$$))$ $>>$ 1$)$ $- 1$$)$
    \State $int\_basis[i]$ =  $w$
    \State $smallest$ = $w$
  \EndFor
  
\EndProcedure

\end{algorithmic}

\end{algorithm*}

We start with a one-dimensional lattice quantum system described by the size of the grid, the number of particles present in the system and a given Hamiltonian operator, 
such as the one described by Equation \eqref{eq: Hamiltonian}. 
%
%
The successive step is to create a representation of the Hilbert subspace, which can be done using integer-binary operations on the computer as explained previously. 
The method we use for this is through a lexicographic computation of next bit permutations, this provides a fast way of computing all the possible combinations in a sorted manner, therefore avoiding the 
requirement of sorting algorithms for later lookup routines. 

The matrix representation and construction of the Hamiltonian operator rests on this basis. This requires to apply the Hamiltonian operator to each of the states in the basis to get the 
matrix elements connecting  the states; this translates into a sparse, Hermitian matrix that is then used for the time evolution of the given system. A sparse storage format is required in this stage, 
given that the expected sizes of the system subspace are very large. 
The initial states are represented as vectors, with a parallel distribution that needs to be consistent with the distribution approach used for the Hamiltonian matrix.

When it comes to the study of disordered systems, many layers of disorder can be introduced to the system in the form of randomness. One way to introduce disorder to the system is to introduce randomness to the 
parameters of the Hamiltonian for example, or to use a random initial state for each simulation \cite{tavora2}.

For the unitary time evolution of the system, there is no requirement to construct the Hamiltonian matrix for each time step. One can just adjust the time parameter and evolve the system using the same 
Hamiltonian representation. This is done by means of the Krylov subspace methods in order to obtain an evolved state: once this time-evolved wavefunction is obtained 
expectation values of any physical observable may be computed as a function of time.

Figure \ref{fig:design} shows a brief description of the workflow implemented.

{\em External libraries and dependencies.} --- The developed application \cite{git} relies on external libraries, namely: \texttt{C++ Boost ver 1.61.0} \cite{boost}, \texttt{PETSc ver 3.7.3} \cite{PETScMan} for 
parallel matrix and vector objects and \texttt{SLEPc ver 3.7.2} \cite{slepc} for time evolution routines. \texttt{PETSc} was compiled using Intel MPI and Intel MKL \cite{mkl} and built to support 
complex datatypes, 64-bit integers and indices, \texttt{FORTRAN} kernels and interfaces.

{\em Basic parallelization.} --- For a fully parallel computation and distribution of objects across processing elements, a row distribution can be used to handle vector and matrix objects. 
Algorithm \ref{Distribution} shows a way in which this can be accomplished. The benefit of using this simple distribution is consistency in relation to external libraries (\texttt{PETSc}). 
The variables \texttt{mpisize\_} and \texttt{mpirank\_} are queried by means of MPI functionality and the parameters are used to decide the row sections of vector and matrix objects that each processing element 
will allocate and have access to. The variables \texttt{nlocal}, \texttt{start} and \texttt{end} refer, respectively,  to the local number of rows and global row indices of each processing element. 
This distribution holds for both the construction of the Hamiltonian and the time evolution procedures, while the construction of the basis rests on different distributions as discussed in the following 
sections.  

\begin{figure*}[htb]
\centering
\includegraphics[scale=0.24]{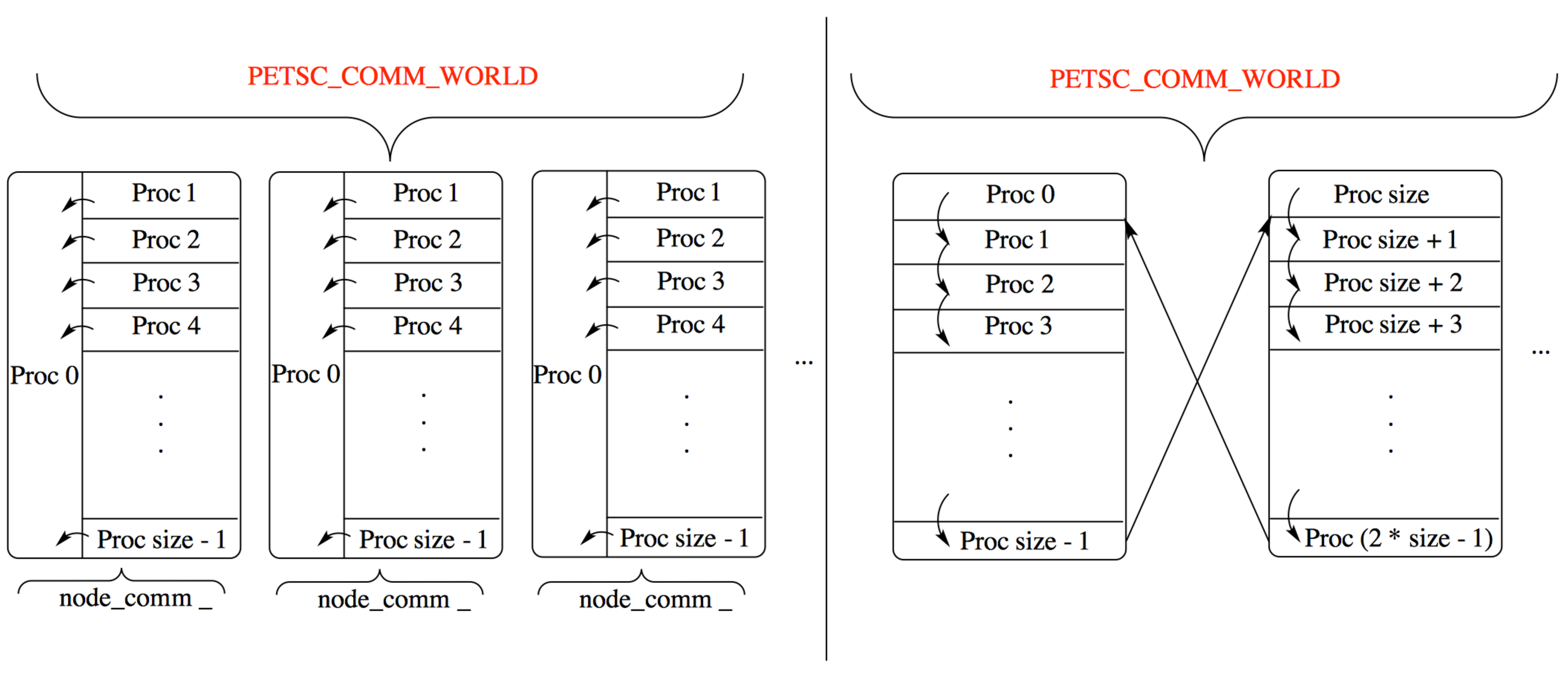}
\caption{Visual representation of the \texttt{node communicator} (left) and \texttt{ring exchange} (right) parallel distribution arrangements for the Hilbert subspace basis representation. 
Each large rectangle represents a node.}
\label{fig:visual} 
\end{figure*}

{\em Basis representation.} --- A basis representation for the Hilbert subspace needs to be computed, given that the construction of the Hamiltonian operator is done by means of the basis. This requires to 
actually compute the integers that represent each of the states of the subspace. To do this, we used a contiguous section of memory of the required size. One could use standard C++ containers for this, but for 
specific design reasons related to memory management in later stages, we allocate memory using the regular mechanism by means of the \texttt{new} command and use raw pointers to access and modify elements. 
This was implemented as shown in Algorithm \ref{Perm}.

In such a way, the \texttt{int\_basis} container gets filled with each of the possible permutations in a sorted manner. The \texttt{smallest\_int()} function is a very simple routine that computes the smallest 
integer value of the bit representation corresponding for any given $L$ and $N$. \texttt{Boost}'s component, \texttt{dynamic\_bitset}, provides easy to use functionality to create binary objects out of each of 
the elements of this container and moving from a binary to integer representation. This functionality was used for the later construction of the Hamiltonian matrix.

We proceed to sequentially compute the values of the basis by means of bit permutations, in such a way that the next permutation of a given \texttt{bitset} corresponds to the binary combination that provides 
the next integer value when translated into an integer representation. One of the benefits of doing this is not only it's performance, but the fact that the resulting integer representation of the basis is 
\textbf{sorted}.

{\em Hamiltonian matrix construction.} --- Once a representation for the basis has been constructed, the Hamiltonian matrix can be computed in a distributed fashion. Each processing element allocates, 
computes, and holds the row sections of the matrix given by the distribution shown in Algorithm \ref{Distribution}. This can be done by mapping the operations in Equation \eqref{eq: Hamiltonian} or any given 
Hamiltonian operator to bitwise operations and transformations and applying them to each of the states in the basis. A lookup routine is then used to obtain the indices of the corresponding matrix elements. 

In light of the sorted basis as described in previous subsection, we can use an efficient method to look up entries of the memory buffer. 
A very good choice is to use a \textbf{binary lookup} and therefore using an algorithm with complexity $\mathcal{O}(\log{N})$ instead of an element-by-element lookup with complexity $\mathcal{O}(N)$.

We use \texttt{PETSc}'s \texttt{Mat} objects to handle Hamiltonian matrix representations. There are basically three different ways \cite{PETScMan} in which a PETSc \texttt{Mat} object can be constructed:

\begin{enumerate}
	\item{Create an instance of the object without specifying preallocation details}
	\item{Create an instance of the object providing estimated values of sizes for preallocation}
	\item{Create an instance of the object providing the exact amount of elements in the diagonal and off-diagonal portions of the matrix with the parallel subdivision taken into account}
\end{enumerate} 

Out of the three methods, the first one is the simplest but performs the worst. This is because of the {\em overhead} related to dynamically resizing memory sections. The second method performs well if a 
good estimation is provided, which usually requires allocating more memory than what is actually required for the object.

\begin{algorithm*}
\caption{Construction of the basis in the \texttt{node communicator} and \texttt{ring exchange} arrangements}\label{BasisCommNode}

\begin{algorithmic}[1]

\Procedure{ConstructIntBasis}{LLInt *$int\_basis$, PetscInt $nlocal$, PetscInt $start$}

  \State LLInt $w$ \Comment{Next permutation of bits}
  \State LLInt $first$ = $first\_int(nlocal,start)$ \Comment{Smallest int of the section of the basis}
  \State $int\_basis[0]$ = $first$
  \For{LLInt $i = 1$ to $nlocal - 1$}
    \State LLInt $t$ = $($first $|$ $($first $- 1$$))$ $+ 1$
    \State $w$ = $t$ $|$ $(((($t \& -t / $($first \& -first$))$ $>>$ 1$)$ $- 1$$)$
    \State $int\_basis[i]$ = $w$
    \State $first$ = $w$
  \EndFor
  
\EndProcedure

\end{algorithmic}

\end{algorithm*}

Our goal is to optimize memory consumption, so the third method described above is the best for our purposes. This requires implementing another routine similar to the one used to find the Hamiltonian matrix 
elements: this computes the elements that each processing element contains in its own section of the matrix, so that a \textbf{preallocation} step can be performed. This can also be performed in parallel in 
order to avoid compromising scalability. 

The procedure introduces computational overhead, but the mechanism performs well enough so that we can use this in order to reduce memory consumption to a minimum, as shown in the Performance section of this 
paper.

{\em Time evolution.} --- After constructing the matrix representation of the Hamiltonian operator in a distributed fashion and consistent with the parallel distribution required for the PETSc and SLEPc 
routines, one can proceed to evaluate the time evolution of the system with an initial state. The initial state can be constructed in many different ways to study different behaviors of the system and should be 
represented as a vector distributed in parallel among processing elements, this can be easily done using PETSc. \cite{PETScWeb}
For the time evolution procedure we can use SLEPc's routine related to the MFN (Matrix Function) component, which provides all the necessary framework with enough versatility to carry out the computation \cite{slepc}.

{\em Basis replication.} --- Table I presents a numerical calculation of memory consumption of the basis for given system sizes. 
A common issue encountered in these parallel design settings is that of basis replication.
Basis replication in this context means having a memory section devoted to contain the integer values representing each of the states in the subspace per each computing element. 
It can be seen from Table I that for a large set of problem sizes basis replication isn't really a problem to be concerned with memory-wise; 
in particular for $L=28$ at half filling and smaller values of $L$, including all the systems with subspace dimension smaller than this, the basis replication poses no problem.
In {\em Message Passing Interface} terms, each MPI process allocates and has access to the memory address of the entire basis.

\textbf{Table I}. Basis memory consumption for different system sizes at half-filling.
\begin{center}
     \begin{tabulary}{1\textwidth}{ C C C }
     \hline
     System sizes & $\mathcal{D}$ & Basis memory (GB) \\
     \hline
    	$L=28$, $N=14$ & $4.01 \times 10^7$ & 0.320 \\
    	$L=30$, $N=15$ & $1.55 \times 10^8$ & 1.25 \\
    	$L=32$, $N=16$ & $6.01 \times 10^8$ & 4.8 \\
	$L=34$, $N=17$ & $2.33 \times 10^9$ & 18.7 \\
    	$L=36$, $N=18$ & $9.08 \times 10^9$ & 72.6 \\
     	$L=38$, $N=19$ & $3.53 \times 10^{10}$ & 282.8 \\
     \hline
\end{tabulary}
\end{center}

Basis replication {\em has} to be avoided if one is interested in evaluating the dynamics of large systems, otherwise given the exponential increase of the size of the Hilbert subspace the basis will quickly 
exhaust local memory resources. This implies that basis {\em distribution} across processing elements has to be implemented to carry out simulations of large systems. 
In the next subsection we present two different distribution arrangements and MPI communication patterns that can be used to overcome this problem. 

\begin{algorithm*}[htb]
\caption{Node communication pattern}\label{NodeComm}

\begin{algorithmic}[1]

\Procedure{NodeComm}{...}

  \State // Communication to rank 0 of every node to find size
  \If{$node\_rank\_$}
    \State $MPI\_Send$(...)
  \Else
    \For{$i=1$ to $node\_size\_$}
      \State $MPI\_Recv$(...)
    \EndFor
  \EndIf
  
  \State // Communication to rank 0 of node to find missing indices
  \If{$node\_rank\_$}
    \State $MPI\_Send$(...missing info...) \Comment{Send unfound information}
    \State $MPI\_Recv$(...required info...) \Comment{Required to finish the construction}
  \Else \Comment{Rank 0 of every node}
    \For{$i=1$ to $node\_size\_$}
      \State $MPI\_Recv$(...missing info of rank $i$)
      \State (...binary search...)
      \State $MPI\_Send$(...required info to rank $i$)
    \EndFor
  \EndIf

  \State (...complete construction with the updated data...)

\EndProcedure

\end{algorithmic}

\end{algorithm*}

\subsection{Parallel distribution of the Hilbert subspace basis representation}

\texttt{Node communicator} {\em approach.} --- We focus our attention now on the first method that was used to overcome the basis replication problem. 
The approach consists of distributing the basis among all the processing elements, except for the first MPI process of each node, which allocates and holds the memory addresses of the entire basis. 
In this scenario, the entire memory required for the basis alone would be: 1 entire basis per computing node plus 1 entire basis distributed across the rest of the MPI processes. 
Computations required to construct the Hamiltonian matrix then require {\em intranode} communications to find missing information. One of the benefits of this solution is the fact that the 
communication is being done inside the node, most MPI implementations benefit from this using hardware locality directives. 
The left panel of Figure \ref{fig:visual} shows a visual representation of this arrangement. 

We used this particular distribution for the construction of the Hamiltonian: however, the time evolution computation should use a global distribution by means of the global communicator 
(\texttt{PETSC\_COMM\_WORLD}) since that provides the best balance and compatibility with PETSc functionality. 

The arrangement can be achieved by enabling a second MPI communicator. We called this second communicator \texttt{node\_comm\_}. As of the release of the MPI 3.0 standard there's a very natural way to 
accomplish this task by means of the MPI shared regions:

\begin{lstlisting}
...
MPI_Comm node_comm_;
MPI_Comm_split_type(PETSC_COMM_WORLD, 
  MPI_COMM_TYPE_SHARED, 
    mpirank_, MPI_INFO_NULL, &node_comm_);
...
\end{lstlisting}

The \texttt{node\_comm\_} MPI communicator can be used to establish communication patterns of computing elements within a local node. Given that the first MPI process of each node contains the entire basis, 
communication can be performed in order to construct the matrix representation of the Hamiltonian operator using the data contained by this computing element.

This new distribution implies that the construction and computation of the basis, allocation details, and Hamiltonian matrix have to be done consistently with the new arrangement. 
In particular this involves a careful indexing of memory locations that correspond to each of the elements of these objects. 
Algorithm \ref{BasisCommNode} shows the approach used to compute the basis for both the \texttt{node communicator} and \texttt{ring exchange} approaches. 

\begin{algorithm*}[htb]
\caption{Ring communication pattern}\label{RingComm}

\begin{algorithmic}[1]

\Procedure{RingComm}{...}

  \State // Collective communication of global indices
  \State $gather\_nonlocal\_values\_$(...);
  \State // Even when rest != 0, Proc 0 always gets the larger section
  \State // of the distribution, so we use this value for the ring exchange buffers
  \State $broadcast\_size\_of\_buffers\_$(...); \Comment{From 0 to all}
  \State (...allocate ring exchange buffers, initialized to $basis$ in a sorted fashion...)
  \State PetscMPIInt $next$ = $($ $mpirank\_$ + 1 $)$ \% $mpisize\_$;
  \State PetscMPIInt $prev$ = $($ $mpirank\_$ + $mpisize\_$ - 1 $)$ \% $mpisize\_$;

  \For{PetscMPIInt $exc=0$ to $mpisize\_ - 2$}
    \State $MPI\_Sendrecv\_replace\_$(...); \Comment{Basis exchange using $prev$ and $next$}
    \State // $source$ is required to find global indices from $nonlocal$ values
    \State PetscMPIInt $source$ = $mod\_$(($prev$ - $exc$), $mpisize\_$);
    \State (...binary lookup of index elements, if found, assign and multiply by $-1$)...
  \EndFor
  
  \State (...flip all the signs: multiply by -1 all found indices of matrix elements...)

\EndProcedure

\end{algorithmic}

\end{algorithm*}

It can be noticed that each MPI process will compute only its own section of the basis. The \texttt{int\_basis} buffer has size $nlocal$\footnote{With the exception of the first MPI process of every node in the \texttt{node communicator} approach, for which $nlocal$ = $basis\_size\_$} given by the distribution in Algorithm \ref{Distribution}. The method \texttt{first\_int()} is a very simple routine that computes the first element in the basis for a given MPI process. 

For the computation of the allocation details and the Hamiltonian matrix the indexing is changed accordingly. In this particular scenario, each MPI process will take ownership of a given subsection of rows in 
the matrix and compute the elements of the matrix by means of its own basis section. Some elements of the matrix are not computed given that the local basis is incomplete, 
so this information is stored and communicated to the first MPI process of the local node in order to complete the procedure by means of the second MPI communicator. 
Algorithm \ref{NodeComm} shows the approach for this particular section of the computation.

\texttt{Ring exchange} {\em approach.} --- Depending on the available memory resources per node of the computational environment in which the application is executed, according to the estimates shown in 
Table I, allocating and constructing one entire basis per node can exhaust the memory resources of the system for large system sizes. 
Even if this could be done, there are still memory resources required for the actual Hamiltonian matrix and time evolution procedure. 
The node communicator version provides a good solution for a large range of system sizes; however for systems that have a very large subspace dimension a fully distributed approach needs to be devised. 

With the \texttt{ring exchange} approach, a full distribution of the basis across all the processing elements is achieved. Communication here is performed in order to exchange sections of the basis and not the 
{\em unfound} elements of the Hamiltonian matrix (this is an important difference as computational load is more balanced in this setup). 
Moreover, the approach can be implemented by means of a single MPI communicator. The right panel of Figure \ref{fig:visual} shows a visual representation of the arrangement.

In this new setup then, each processing element exchanges sections of the basis in order to compute the elements of the Hamiltonian matrix. After $mpi\_processes - 1$ exchanges, all the elements have been 
computed and the Hamiltonian matrix is computed and distributed. 
With this setup, a {\em linear scaling} in memory is achieved which means that with increasing number of processing elements the amount of memory per MPI process required decreases linearly. 
However, {\em time scaling} is compromised, as increasing the number of processing elements will require more communication steps. 
Therefore, we expect that with this setup the time required to compute the Hamiltonian matrix will increase in relation to previous procedures. 

The basis can be computed and distributed using Algorithm \ref{BasisCommNode} using the corresponding global indices, while the computation of allocation details and the Hamiltonian matrix the indexing has to 
be changed accordingly. The major difference from the \texttt{node communicator} approach is the communication step. 

Algorithm \ref{RingComm} shows the procedure used to perform the ring exchange of the basis in order to compute the Hamiltonian matrix. The first step is to to perform a collective communication so that every 
processing element holds the $nonlocal$ values of the global indices of each subsection of basis being communicated, in total there are \texttt{mpisize\_} nonlocal global indices to be stored per computing element. This has to be done to keep track of the global positions of the elements in the Hamiltonian matrix. 
Afterwards, since the number of processors is a generic parameter, different MPI processes may hold larger sections of the basis than others. 
To account for this, we use the largest local size to allocate the exchanging buffers, initialized to the elements of the basis with the remaining elements set to zero, as this will not affect the binary lookup given that no state in the basis is represented by the 
zero value. This case is important when the \texttt{rest} is different than zero (see Algorithm \ref{Distribution}). These exchanging buffers have to be additionally allocated in memory and their size is \texttt{nlocal} of the first process in the communicator set, as this processing element will hold the largest local basis size when \texttt{rest} is different than zero. 

The variable \texttt{source} identifies the MPI process that sends the section of the basis; this is required in order to align the indices with global values. 
The last step is to perform the ring exchange and the binary lookup procedure until all sections of the basis have been explored. 
In order to keep track of the already found index elements, we set them to negative values and when the procedure is done we flip them back to positive entries. 
At the end of the procedure all the index elements of the Hamiltonian matrix have been found and computed. 

\begin{figure*}[ht!]
\fontsize{14}{10}\selectfont 
\centering
\includegraphics[scale=1.0]{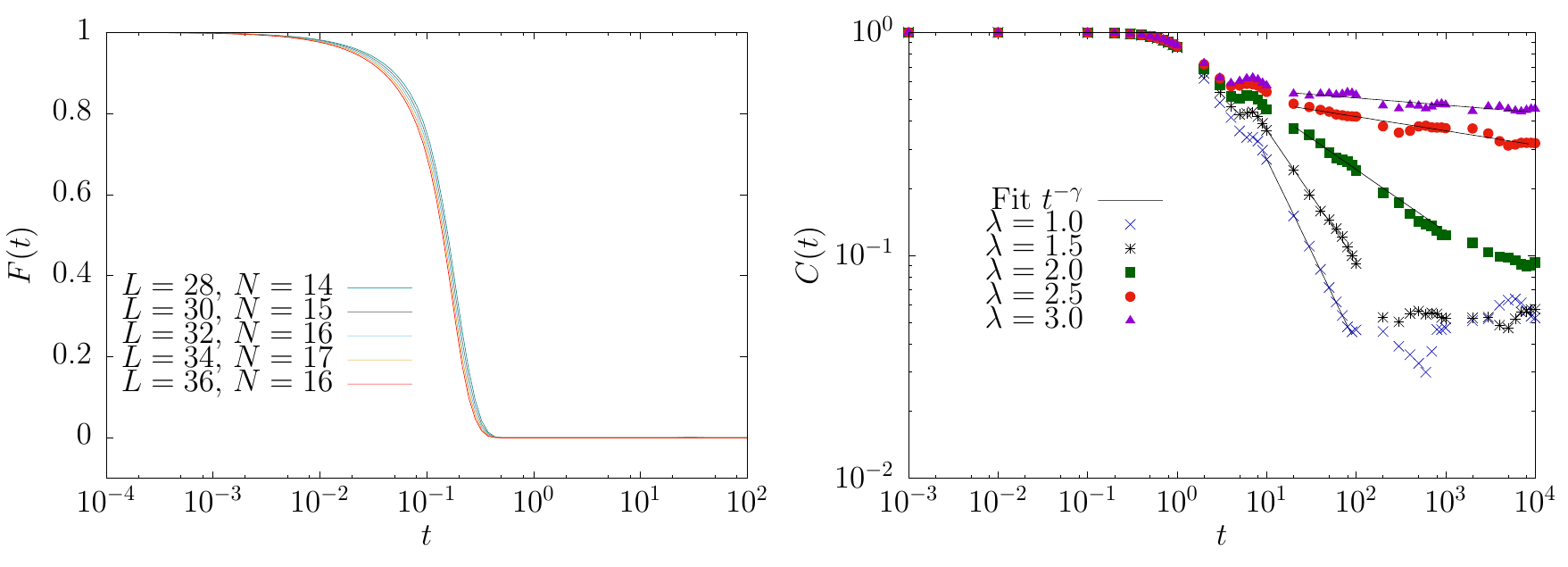}
\caption{Survival probability $F(t)$ for different system sizes on a system with no disorder (left) and temporal autocorrelation function for the Aubry-Andr\'e model non-interacting case with 1 particle 
($L = 55$) using periodic boundary conditions for different values of $\lambda$ (right).}
\label{fig:loschmidt} 
\end{figure*}

\section{Results}

With this implementation, we may now compute the unitary time evolution described by Equation \eqref{eq: Schrodinger} of a many-body quantum system.

{\em Survival probability.} --- A simple yet physically relevant dynamical observable is the survival probability, which is defined as follows.
We start with the system prepared with a given initial state $\ket{\Psi(0)}$ at $t=0$. 
The probability of finding the system in state $\ket{\Psi(0)}$ at time $t$ is the so-called survival probability given by 

\begin{equation}
\label{eq: SP}
F(t) = |A(t)|^2 \equiv |\bra{\Psi(0)}e^{-iHt}\ket{\Psi(0)}|^2
\end{equation}

where $A(t)$ is the survival amplitude. We evaluated this quantity by means of the Krylov subspace in systems of Hilbert space dimension of over 9 billion. 
The survival probability can be shown to be the Fourier transform of the local density of states (LDOS) corresponding to the initial state chosen: for an ergodically filled LDOS 
(expected for generic thermalizing or chaotic systems) the power-law decay of $F(t)$ is enhanced whereas this power-law can be suppressed due to multifractality or build-up of correlations among the 
eigenstates of the Hamiltonian. 
We employ the rate of power-law decay as a measure of the chaoticity or lack thereof in the systems \cite{tavora1, tavora2}.

{\em Clean systems.} --- We first consider a clean hardcore bosonic chain with $h=0$ in Equation \eqref{eq: Hamiltonian}. Due to its integrability and lack of chaoticity the power-law is expected to be suppressed.
Figure \ref{fig:loschmidt} exposes a very clean behavior of the survival probability given that Equation \eqref{eq: SP} doesn't introduce any form of disorder to the system; 
for these simulations we used $t=1$ and a weak interaction of $V=0.2$, which is very close to the free fermionic system. 
The initial state used in these simulations is the N\'eel ordered state, which in our binary representation is the state given by (...010101). 
We find the power-law decay to be $\gamma \approx 0.97$ consistent with the fact that the model is integrable and non-ergodic.\\

{\em Quasidisordered systems.} --- We evaluated the survival probability of a quasidisordered system, namely the interacting and noninteracting Aubry-Andr\'e model with the Hamiltonian operator given by 
Equation \eqref{eq: Hamiltonian}, with $h \neq 0$; we used with periodic boundary conditions and gathered disorder averaged results over an ensemble of random initial states. 
To produce cleaner results we measured a different form of the survival probability: the so-called {\em temporal autocorrelation function} which is given by \cite{ketzmerick}

\begin{align}
C(t) = \frac{1}{t}\int_{0}^{t}|\bra{\Psi(0)}\ket{\Psi(t^{\prime})}|^{2}dt^{\prime}
\end{align}

The system is usually studied as a function of the parameters $t$, $V$ and $h$ in Equation \eqref{eq: Hamiltonian}. We define $\lambda \equiv h / t$ and vary this parameter for a fixed value of $V$. Figure \ref{fig:loschmidt} 
shows the behavior of the temporal autocorrelation function for five different regimes in the non-interacting case for a single particle: 
$\lambda<2$ corresponds to the behavior of the delocalized or extended states, $\lambda>2$ gives rise to the localized state behavior, and $\lambda=2$ is known to be the critical point, 
corresponding to neither localized nor extended states \cite{ketzmerick}. 
Indeed as shown earlier \cite{ketzmerick} here we confirm that in this noninteracting setting the power-laws of the temporal decay are quite small due to multifractality in the eigenspectrum. 
$\gamma$ ranges from $0-0.83$ as one transits from the localized to delocalized phases, with $\gamma \approx 0.27$ at the critical point.

An interesting case of study is to evaluate the temporal evolution of the system described with the Aubry-Andr\'e model for the interacting case, by varying the particle density. 
We proceed in a similar fashion as we did for the single particle case and study the dynamics of the system using increasing densities $\rho \equiv N / L$.

\begin{figure*}[ht!]
\centering
\includegraphics[scale=1.0]{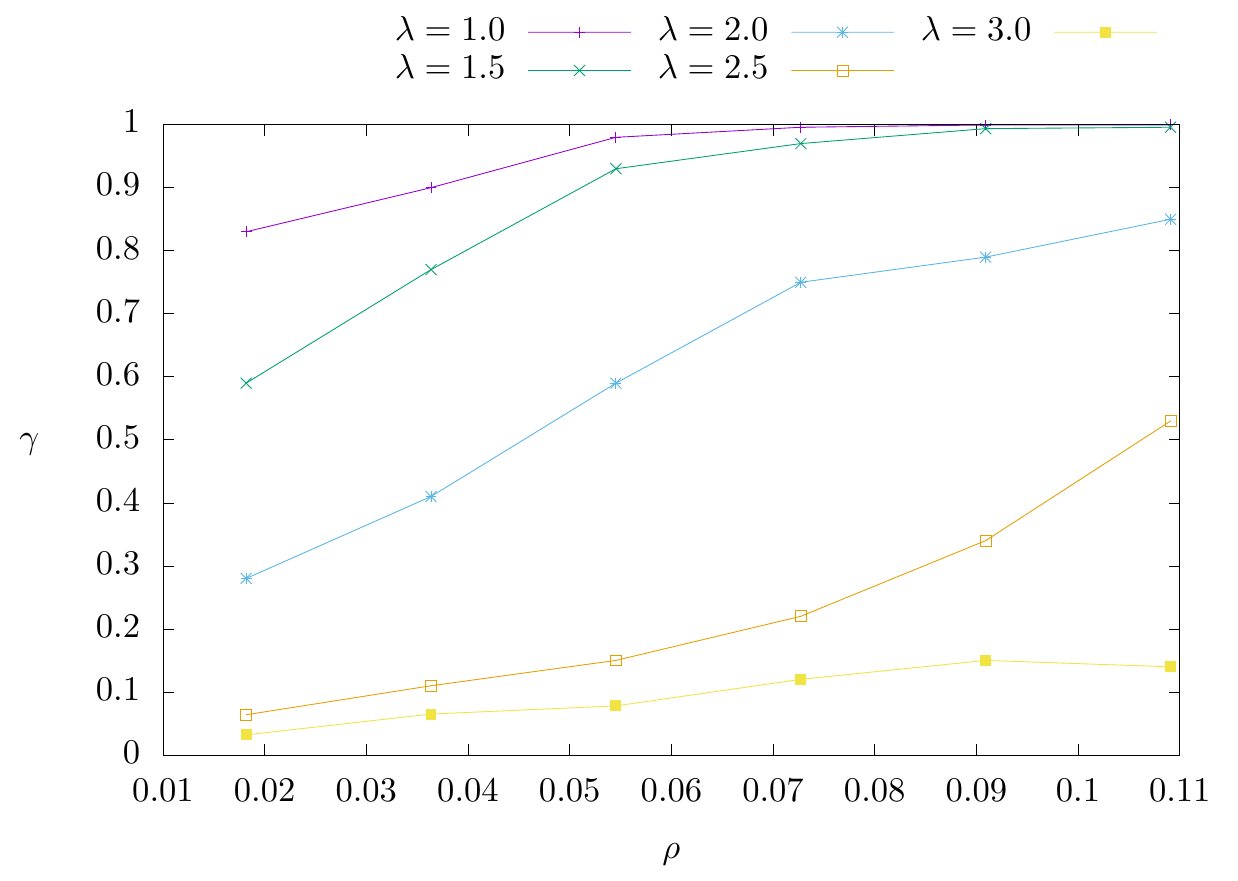}
\caption{Temporal decay exponent $\gamma$ for different densities ($\rho \equiv N / L$) and values of $\lambda$ for the interacting Aubry-Andr\'e model with periodic boundary conditions and $L=55$.}
\label{fig:variation} 
\end{figure*}

Figure \ref{fig:variation} shows the variation of the temporal decay exponent as a function of both $\lambda$ and the density $\rho$; it can be seen that increasing the density of the system favors the 
decay of the survival probability. This may be interpreted to mean \cite{tavora2} that increasing the density favors thermalization. Although according to Ref. \cite{tavora2} 
the regime for which $0<\gamma<1$ no thermalization occurs at all. However we believe that for this system, because the presence of multifractality in the noninteracting limit already lead to slow 
temporal decay and $\gamma < 1$ even in the delocalized phase $\lambda < 2$ (as seen in the Fig. \ref{fig:variation} and previously shown in Ref. \cite{ketzmerick}) 
interactions cannot give $\gamma > 2$ required for ergodic filling of LDOS.

It can also be seen from our analysis that $\gamma \leq 1$ for the $\lambda$ values considered. 
This indicates that, unlike the truly disordered case \cite{tavora1,tavora2}, faster fully ergodic decay is not present at these densities presumably due to the multifractality present in the single particle limit. 
However for weak quasidisorder strength $\lambda < 2$, the exponent $\gamma \rightarrow 1$; this indicates the emergence of chaotic initial states with decreasing correlations among eigenstates.
For $\lambda > 2$ the presence of non-chaotic states with significant correlations among the eigenstates leads to a very slow decay, giving $\gamma < 1$. 
In either case multifractality present in the noninteracting limit exacerbates the slow decay.


%



\begin{figure*}[ht!]
\fontsize{14}{10}\selectfont 
\centering
\includegraphics[scale=1.0]{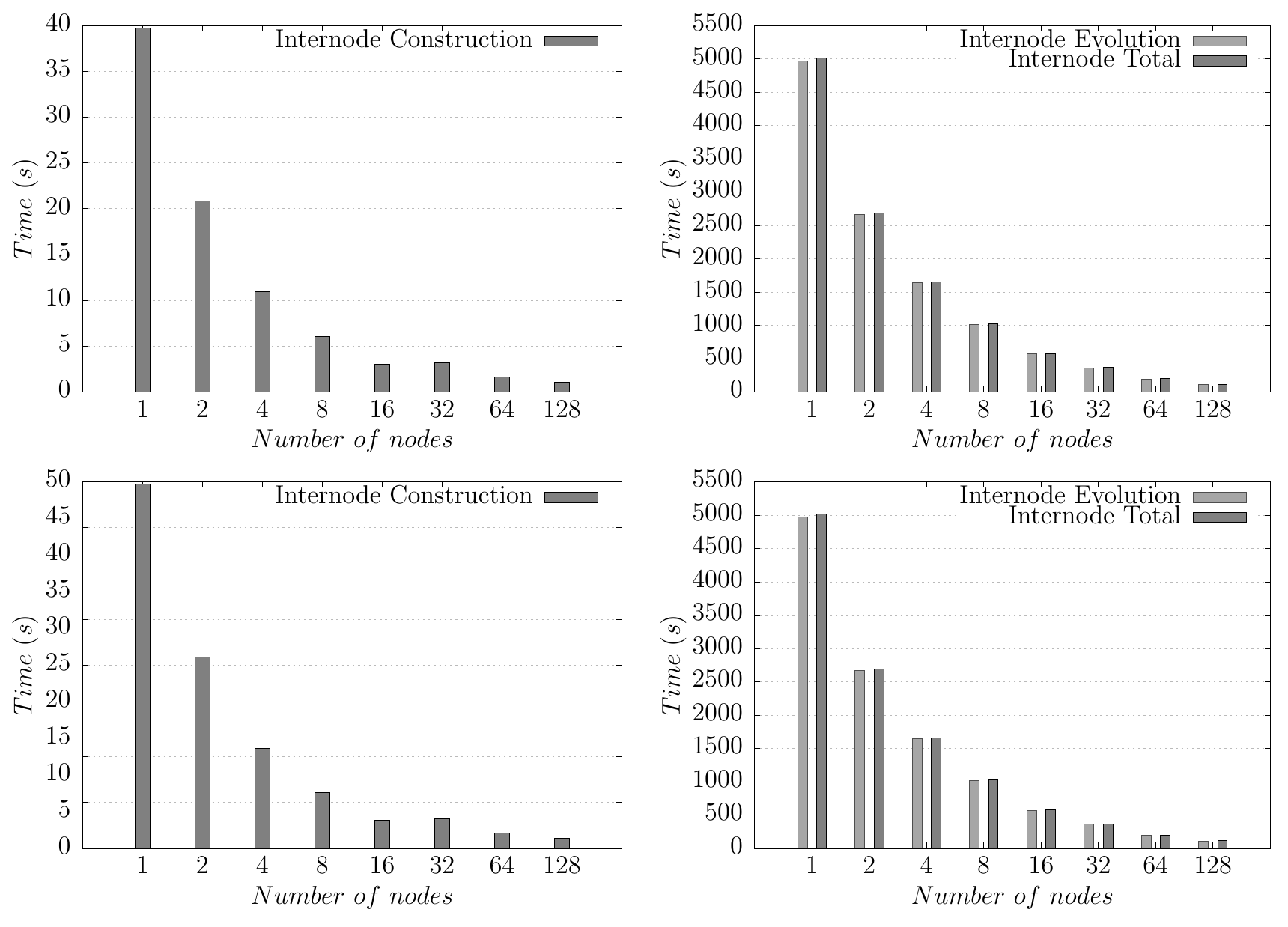}
\caption{\small{Strong scaling using a system with $L=28$ at half-filling with a subspace dimension of 40,116,600, evolved up to $t=100$ with a tolerance of $10^{-7}$ for varying number of computing nodes: (top) 
using a replicated basis approach, (bottom) using the \texttt{node communicator} approach, (left) time required for allocation and computation of the Hamiltonian matrix and (right) time required for 
time-evolution procedure and overall wall-time}.}
\label{fig:bars} 
\end{figure*}

\section{Performance analysis}

{\em System.} --- Table II shows the technical description of the system in which the application was tested and used for production runs.

{\em Compilers and libraries.} --- The application has been compiled and tested using the C++ MPI wrappers of the Intel MPI compiler version 5.1.3. For underlying linear algebra operations we use Intel MKL 
version 11.3.3. The versions of the higher-end libraries used are: Boost v1.61.0, PETSc v3.7.3, SLEPc v3.7.2. 

{\em Performance results.} --- We started by evaluating the performance of the first instance of the application in a parallel environment. As was described in the Implementation section the first approach uses 
basis replication among all processing elements to construct the matrix representation of the Hamiltonian matrix, however, the computation and distribution of this object and the later time evolution step 
are performed in a fully distributed fashion.

\textbf{Table II.} Technical description of platform where simulations were executed.
\begin{center}
     \begin{tabulary}{1\textwidth}{ p{0.2\textwidth} C }
     \hline
     Description & Galileo-CINECA\\
     \hline
     Model & IBM NeXtScale\\
     Architecture & Linux IB Cluster\\
     Nodes & 516\\
     Processors & 2 x 8-cores Intel Haswell\\
     RAM & 128 GB/node\\
     Internal Network & IB with 4x QDR switches\\
     Total on-board memory (RAM) & 66 TB\\
     \hline
\end{tabulary}
\end{center}

Figure \ref{fig:bars} shows the strong scaling behavior of the application in the two most important steps: the construction of the Hamiltonian object and the time evolution. As can be observed from 
Figure \ref{fig:bars}, the rest of the procedures such as computing the basis and initializing routines perform very well and account for only a negligible portion of the computation time. 

It can be seen that the time for the construction of the Hamiltonian object is very small compared to the actual time evolution procedure even for a relatively small physical time parameter ($t=100$). 
Indeed one of our objectives was to keep the wall time required to construct the Hamiltonian object small when compared to the time evolution, 
in order to avoid compromising scalability for a very large number of processing elements. Our results show a good scaling behavior even for large amount of processing elements 
(2048 MPI processes mapped to each core, as shown in Figure \ref{fig:bars}). An MPI approach was also required, given that for larger problem sizes ($L=30$ at half filling and larger, for instance) 
cannot be solved by means of a single computational node because of memory requirements and to enable the usage of more than one computational node for calculations.

Another important aspect is that constructing and solving the system for the given parameters takes under 2 minutes overall with 128 nodes on the Galileo-CINECA machine. 
This is an important factor if, for instance, one is interested to undertake disorder averages over random initial states or another random parameter of study.

\texttt{Node communicator} {\em approach.} --- As was described in the previous section, basis replication is a major issue when it comes to large problem sizes. We developed the \texttt{node communicator} 
version with two objectives in mind: avoiding basis replication and therefore opening up the possibility of studying larger system sizes, and simultaneously retaining strong scaling behavior in the process. 
With this approach we achieved both as shown in the bottom panel of Figure \ref{fig:bars}.

\begin{figure*}[ht!]
\centering
\includegraphics[scale=1.0]{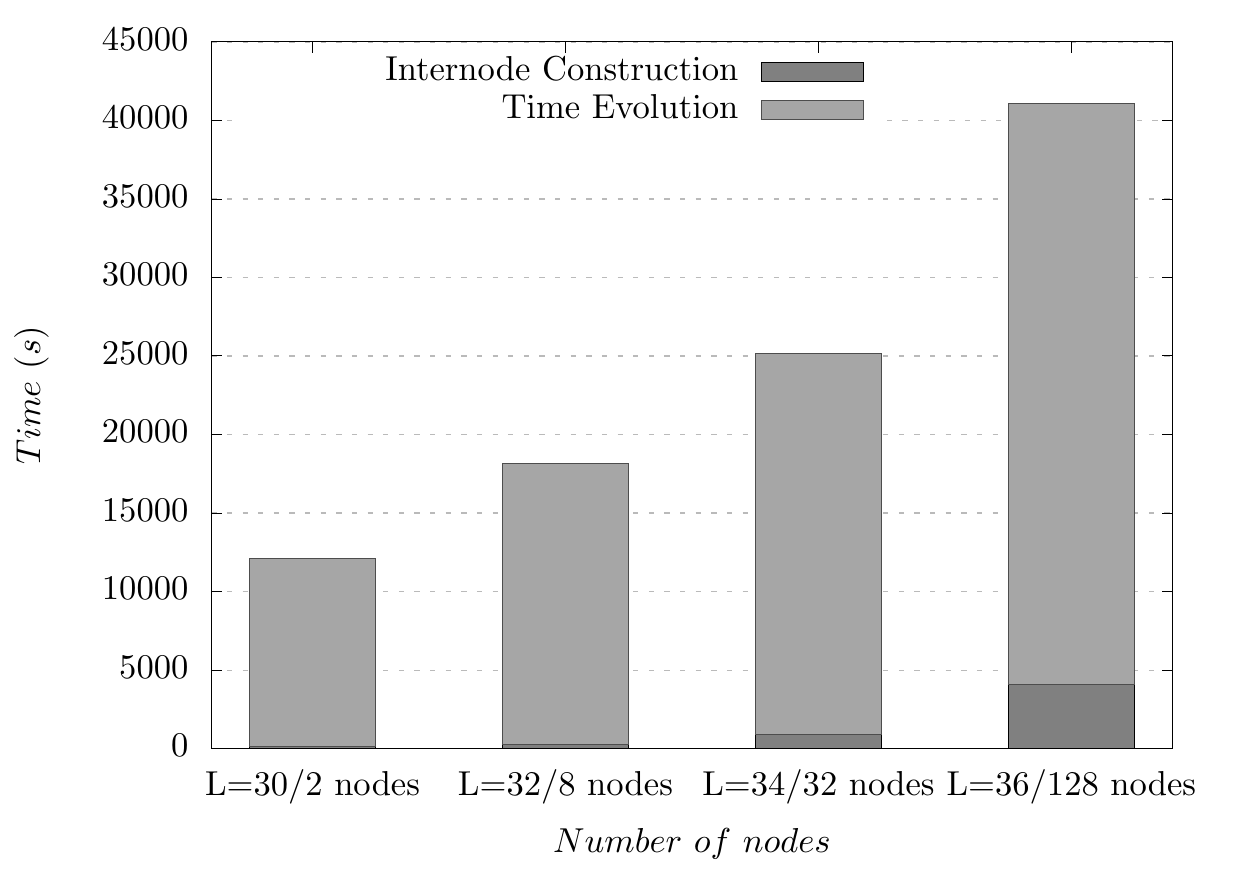}
\caption{Running time of the application for the construction of Hamiltonian and time evolution using different problem sizes at half-filling, evolved up to $t=100$ with a tolerance of $10^{-7}$ for varying 
number of computing nodes using the \texttt{Ring exchange} approach.}
\label{fig:ring} 
\end{figure*}

On the Galileo-CINECA machine we managed to solve a system with size up to $L=34$ at half filling (2,333,606,220 states) using this approach, 
although this version is also very useful on machines with lesser memory per node given that full replication of the basis is avoided, and actual computation and scalability of the overall application are not 
compromised. 
The actual time evolution of the system is unchanged from the replicated basis approach as seen in top panel of Figure \ref{fig:bars} showing the same behavior as the previous version. 
In Figure \ref{fig:bars} it can also be observed that the increase in time required to construct the Hamiltonian operator is very small compared to the time evolution of the system even for this relatively 
big system size, which was exactly our goal. Other different approaches, such as a systematic re-computation of sections of the basis to avoid storing this object in memory, proved to be inefficient. 

Strong scaling behavior for this approach is not linear, but the results obtained are satisfactory when it comes to actually solving the problem using a large number of processing elements. 
A {\em flat} behavior can be seen at the transition from 16 to 32 nodes, which we believe could be related to the network arrangement of the machine so that communication beyond 16 nodes could be using different 
switches in the interface (higher tree in the network arrangement). 
This behavior is present in both of the approaches used; so we can conclude it is unrelated to the parallel distribution, hence requiring no further investigation of this aspect. 

\texttt{Ring exchange} {\em approach.} --- The ring exchange version was developed with the purpose to open the possibility to study even larger problem sizes. 
Recalling the description in the Implementation section, in this version no memory replication occurs in the application whatsoever. 
Memory requirements decrease linearly with increasing number of computing nodes, but {\em time scalability} is indeed compromised given that a larger number of processing elements implies more overall 
communications. So in perspective, a strong scaling view of the performance like the one shown for the previous developments is not a meaningful for this approach.

With this version, our goal was to have the ability to study even larger systems sizes by means of a full distribution of memory and computation overall. 
To demonstrate a perspective of the performance accomplished by this approach we show the time of computation required in the most important sections of the application with increasing problem size and 
processing elements in Figure \ref{fig:ring}.

It can be seen that the time to construct the Hamiltonian matrix object becomes important as the problem size increases, up to roughly 10\% of the overall computation time for the largest case ($L=36$ at 
half-filling). 
With this approach, most of the communication done during the construction of the Hamiltonian object occurs inside the node which is beneficial in general terms, but when using a large number of processing 
elements the communication steps start to account for a considerable percentage of the overall execution time. Even so, with this approach we were able to solve the $L=36$ at half-filling system; a system that has a Hilbert subspace dimension of more than 9 billion elements.\\

{\em Overall memory consumption.} --- With the approaches described in this paper we have managed to overcome the problem of basis replication by means of MPI distribution and communication patterns; 
however, a practical limit set by the size of the actual Hamiltonian matrix is still present. 

Table III shows an estimation of the memory required to store the matrix representation of the Hamiltonian operator and the overall memory occupation of the application. 

The methodology of Krylov subspaces requires this matrix representation to be stored in memory to evaluate the projections on the subspace. Such objects are also required to be stored internally by the 
library (\texttt{SLEPc}). The overall memory occupation presented in Table III during the time evolution procedure does not include the memory required by the basis given that the basis object can be 
deallocated before this routine. We have measured the full occupation using \texttt{PETSc}'s internal profiler \cite{PETScMan}.
Therefore we are confident that our design would have nicely scaled in its performance for even larger system sizes ($L \geq 38$) were more memory resources made available.

\setcounter{footnote}{4}

\textbf{Table III}. Hamiltonian matrix and overall memory occupation at half-filling.
\begin{center}
     \begin{tabulary}{1\textwidth}{ C C p{0.17\linewidth} p{0.18\linewidth} }
     \hline
     System sizes & $\mathcal{D}$ & Matrix memory\footnotemark (GB) & Full occupation (TB)\\
     \hline
     $L=28$ & $4.01 \times 10^{7}$ & 18 & 0.053 \\
     $L=30$ & $1.55 \times 10^{8}$ & 75 & 0.220 \\
     $L=32$ & $6.01 \times 10^{8}$ & 308 & 0.902 \\
     $L=34$ & $2.33 \times 10^{9}$ & 1269 & 3.72 \\
     $L=36$ & $9.08 \times 10^{9}$ & 5 227 & 15.3 \\
     $L=38$ & $3.53 \times 10^{10}$ & 21 490 & 63.0 \\
     \hline
\end{tabulary}
\end{center}

\setcounter{footnote}{5}
\footnotemark{Estimation}

\section{Conclusions}

An application to study the dynamics of quantum correlated systems suitable to be executed on massively parallel supercomputers has been developed and tested in this work. 
We have used high-performing libraries in conjunction with distributed memory algorithms in order to study large quantum systems with subspace dimension of over 9 billion states with the computational 
resources available. The main computational algorithm which we implemented and parallelized and optimized was the Krylov subspace technique.

To check the validity of the results we have studied and presented the dynamics of known models \cite{ketzmerick}\cite{tavora2} as well as unexplored regimes in the quasidisordered system. 
The latter are gaining experimental relevance these days in the study of the physics of thermalization and localization. We showed that in these quasidisordered systems the survival probability 
decays as a power-law at long times with a dynamical exponent which is quite small $\gamma < 1$; we attribute this to the presence of multifractality already present in the single-particle system as well.

Different parallelization strategies have been implemented to gradually overcome memory scaling problem: these were the Node communicator approach and the Ring exchange approach.
Their performance was analyzed against that of the naive replicated version, and good scaling was observed with the increase of the number of nodes.

But a practical limit is still present. As can be observed from Table III, evaluating the dynamics of a 
system with $L=38$ at half-filling would required at least around 63 TB of memory distributed across many computational nodes, which corresponds roughly to the entire on-board memory of the machine used to 
run simulations (see Table II). 
Though this natural limit is still present and restricts the possibility to simulate even larger systems, our implementation has overcome the large memory requirements set by the basis to be able to simulate 
extremely large quantum systems. 

{\em Acknowledgements.} --- We thank L. Santos for useful discussions. 

\end{multicols}

\end{document}